*Article*

# Winds Versus Jets in Active Galactic Nuclei

**David Garofalo [1,*], Chandra B. Singh [2], Atticus Magerko [1], Marco Botello [1], Sophia Soto [1] and Max North [3]**

[1] Department of Physics, Kennesaw State University, Marietta, GA 30060, USA; lmagerko@students.kennesaw.edu (A.M.); mbotello@students.kennesaw.edu (M.B.); ssoto6@students.kennesaw.edu (S.S.)
[2] House No. 399, Gauriputra Marg, Town Planning, Kathmandu-32, Nepal; chandratalk@gmail.com
[3] Department of Information Systems, Kennesaw State University, Marietta, GA 30060, USA; mnorth@kennesaw.edu
* Correspondence: dgarofal@kennesaw.edu

**Abstract**

The well-established anti-correlation between disk winds and relativistic jets in X-ray binaries is often interpreted in a scale-invariant black hole accretion context. If so, active galactic nuclei (AGN) should exhibit a direct mass-scaled analog. We test this prediction across FRII radio quasars, radio-quiet quasars, and jetted and non-jetted Narrow Line Seyfert 1 galaxies (NLS1) in spirals, among others. They exclude simple scale invariance. The highest-velocity winds occur exclusively in radio-quiet quasars, while powerful FRII quasars host systematically weaker winds despite equally large black hole masses. Jetted NLS1s show strong wind suppression consistent with X-ray binary behavior, whereas FRII quasars occupy a distinct regime in which jets and winds coexist. Black hole mass and spin magnitude alone cannot account for this dichotomy. We argue that the angular momentum direction of the disk relative to that of the black hole (aligned versus anti-aligned or co-rotation versus counter-rotation) is the critical parameter: secularly fueled spiral systems and most post-merger systems favor co-rotation, which is associated with compact ISCO radii, high radiative efficiency, strong winds, and jet suppression, while the counter-rotating subset of merger-influenced ellipticals can sustain powerful jets alongside moderate winds. Moreover, while spiral AGN and merger-driven radio-quiet quasars experience similar strong jet/wind anti-correlation, they cannot be treated as strict scaled analogs of X-ray binaries, which undergo rapid state transitions involving magnetic flux redistribution absent in AGN. At least two distinct wind–jet regimes therefore operate across the mass scale. We identify the details of this behavior across AGN subclasses.

## 1. Introduction

Accretion onto black holes produces two of the most powerful forms of energy release in the Universe, relativistic jets and disk-driven winds. In stellar-mass black hole X-ray binaries, two decades of observations have established a striking anti-correlation between these outflows—steady jets dominate in hard states while disk winds emerge in soft states, suggesting a fundamentally competitive relationship between the two [1-3]. If black hole accretion is scale invariant, this wind–jet interplay should extend seamlessly to supermassive black holes in active galactic nuclei (AGN). Yet the AGN population presents a more complex picture. While some systems appear to mirror the X-ray binary be-havior, others—particularly powerful radio quasars—do not exhibit the same degree of mutual exclusivity between winds and jets.

Both winds and jets ultimately derive their power from the accretion flow, but they are launched through distinct physical mechanisms. Relativistic jets are generally thought to require strong large-scale magnetic flux threading the black hole (e.g., [4,5], whereas winds originate from the accretion disk through radiative, thermal, or magnetically driven processes (e.g., [1,2,6,7]). Because these outflows depend differently on the structure of the accretion flow, magnetic flux distribution, and disk geometry, many theoretical and observational studies have suggested an inverse relationship between jet and wind production. Whether the wind–jet anti-correlation observed in X-ray binaries extends to supermassive black holes remains an open question.

We examine the wind–jet connection across spiral and merger/elliptical AGN, spanning Seyfert galaxies, radio-quiet quasars, and FRII radio quasars ([8] and references therein; see also [9]) and show that direct scale invariance alone appears insufficient to explain the observed phenomenology. We argue that an additional governing parameter—accretion disk alignment or anti alignment relative to the black hole angular momentum—plays a central role in regulating the competition between winds and jets across the mass scale.

In Section 2 we explore ionized outflows, black hole mass, and black hole spin in various AGN, provide statistics, and show that a simple scale invariance is insufficient to explain the data. We then provide an interpretation, illustrating our idea that AGN obey two separate tracks in their jet-wind interplay, unlike X-ray binaries. In Section 3 we conclude.

## 2. Winds and Jets in AGN

In Table 1 we compile outflow velocities, expressed in units of the speed of light, together with black hole masses for four classes of AGN: FRII radio quasars/broad-line radio galaxies (BLRGs), radio-quiet quasars (RQQs), non-jetted spiral Seyfert galaxies, and jetted narrow-line Seyfert 1 galaxies (NLS1s). The sample includes X-ray ultra-fast outflows (UFOs), warm absorbers, UV absorption outflows, broad absorption line (BAL) winds, and large-scale molecular outflows reported in the literature (e.g., UFOs in radio-loud AGNs; UFOs in AGNs; Warm absorbers in Seyfert galaxies; Powerful molecular outflows in local ULIRGs; Molecular winds in local AGN).

Throughout this work we adopt the working hypothesis that these various outflow phenomena represent different manifestations of accretion-disk-driven winds observed under different physical conditions and spatial scales. In this picture, UFOs originate in the innermost regions of the accretion flow, where intense radiation fields produce highly ionized relativistic outflows (e.g., [10,11]). Warm absorbers and UV narrow absorption outflows arise at larger radii where the gas is less ionized and correspondingly slower (e.g., [12]). BAL winds are widely interpreted as powerful radiatively accelerated disk winds viewed along favorable lines of sight, while molecular outflows are interpreted as the large-scale interaction of AGN-driven winds with the surrounding interstellar medium (e.g., [13,14]).

The physical origin of AGN outflows, particularly X-ray UFOs, remains an active area of research. Proposed driving mechanisms include radiative acceleration, thermal driving, magnetocentrifugal winds, and combinations thereof (e.g., [7,15,16]). In the present work, however, we assume that the observed outflow classes are fundamentally linked to radiation generated by radiatively efficient, sub- or near-Eddington accretion onto the central black hole. Super-Eddington accretion is therefore not considered within our framework. Furthermore, unlike conventional interpretations that treat Blandford–Payne outflows as an independent wind component, we regard the magnetocentrifugal disk outflow as an integral part of the jet-launching system, providing the slower, matter-dominated flow that contributes to jet support and collimation.

Throughout this work, the distinction between radio-loud and radio-quiet AGN is consistent with the conventional radio-loudness criterion of [17], ensuring that the source classifications adopted here are compatible with the standard AGN literature. Our sample consists of 62 AGN divided into four subclasses: 9 radio-loud quasars/broad-line radio galaxies (FRII RLQs/BLRGs), 27 radio-quiet quasars hosted primarily in ellipticals or merger systems, 25 radio-quiet spiral AGN including Seyferts and non-jetted NLS1s, and 1 jetted NLS1. These are plotted in Figure 1. Several key features emerge.

**Table 1.** Objects and references of objects plotted in Figure 1.

| **Radio-Loud Quasars.** | | | | | | |
|---|---|---|---|---|---|---|
| **Source** | $v/c$ | **Wind Ref** | $M_{BH}(M_{\odot})$ | **Mass Ref** | **Spin** | **Spin Ref** |
| 3C 111 | 0.041 | [10] | $3.2 \times 10^8$ | [18] | — | — |
| 3C 390.3 | 0.146 | [10] | $4.0 \times 10^8$ | [18] | — | — |
| 3C 445 | 0.033 | [19] | $1.4 \times 10^8$ | [20] | — | — |
| 4C + 74.26 | 0.012 | [21] | $6.3 \times 10^9$ | [18] | — | — |
| 3C 382 | 0.0033 | [22] | $9.5 \times 10^8$ | [23] | — | — |
| 3C 273 | 0.0122 | [24] | $6.6 \times 10^9$ | [24] | — | — |
| PKS 1549-79 | 0.276 | [25] | $2.5 \times 10^8$ | [18] | — | — |
| 3C 105 | 0.227 | [25] | $4.0 \times 10^8$ | [18] | — | — |
| MG J0414 + 0534 | 0.28 | [25] | $1.0 \times 10^9$ | [26] | — | — |
| **Radio-Quiet Quasars (Ellipticals/Mergers).** | | | | | | |
| **Source** | $v/c$ | **Wind Ref** | $M_{BH}$ | **Mass Ref** | **Spin** | **Spin Ref** |
| Mrk 205 | 0.10 | [10] | $4.0 \times 10^8$ | [18] | — | — |
| PG 1211 + 143 | 0.0577 | [27] | $4.0 \times 10^7$ | [28] | — | — |
| PDS 456 | 0.29 | [29] | $1.6 \times 10^9$ | [30] | — | — |
| IRAS 13349 + 2438 | 0.27 | [31] | $1.0 \times 10^9$ | [32] | $> 0.93$ | [33] |
| APM 08279 + 5255 | 0.40 | [34] | $1.0 \times 10^{10}$ | [35] | — | — |
| SDSS J1441 + 4055 | 0.47 | [26] | $5.0 \times 10^9$ | [26] | — | — |
| SDSS J1029 + 2623 | 0.58 | [26] | $6.3 \times 10^8$ | [26] | — | — |
| SDSS J1529 + 1038 | 0.25 | [26] | $7.9 \times 10^8$ | [26] | — | — |
| SDSS J1353 + 1138 | 0.34 | [26] | $2.5 \times 10^9$ | [26] | — | — |
| SDSS J0904 + 1512 | 0.26 | [26] | $2.0 \times 10^9$ | [26] | — | — |
| Q2237 + 0305 | 0.18 | [26] | $1.31 \times 10^9$ | [26] | — | — |
| SDSS J1128 + 2402 | 0.56 | [26] | $5.0 \times 10^8$ | [26] | — | — |
| SDSS J0921 + 2854 | 0.41 | [26] | $7.9 \times 10^8$ | [26] | — | — |
| HS 1700 + 6416 | 0.38 | [26] | $1.6 \times 10^{10}$ | [26] | — | — |
| PG 1115 + 080 | 0.37 | [26] | $6.3 \times 10^8$ | [26] | — | — |
| HS0810 + 2554 | 0.43 | [26] | $4.0 \times 10^8$ | [26] | — | — |
| PG 1202 + 281 | 0.108 | [36] | $4.0 \times 10^8$ | [36] | — | — |
| 2MASS J165315 + 2349 | 0.11 | [36] | $1.0 \times 10^7$ | [36] | — | — |
| PG 0947 + 396 | 0.305 | [36] | $5.0 \times 10^8$ | [36] | — | — |
| LBQS 1338-0038 | 0.152 | [36] | $6.3 \times 10^7$ | [36] | — | — |
| PG 0804 + 761 | 0.13 | [36] | $2.0 \times 10^8$ | [36] | — | — |
| PG 1114 + 445 | 0.072 | [36] | $6.4 \times 10^8$ | [36] | — | — |
| IRAS F11119 + 3257 | 0.255 | [37] | $2.0 \times 10^8$ | [38] | — | — |
| PG 1001 + 291 | 0.28 | [39] | $1.3 \times 10^8$ | [39] | — | — |
| ARK 120 | 0.269 | [40] | $1.3 \times 10^8$ | [18] | $> 0.85$ | [41] |
| MR 2251-178 | 0.137 | [40] | $1.6 \times 10^8$ | [18] | — | — |
| IRAS 05189-2524 | 0.11 | [42] | $2.5 \times 10^7$ | [18] | — | — |
| **Radio-Quiet AGN (Spirals).** | | | | | | |
| **Source** | $v/c$ | **Wind Ref** | $M_{BH}$ | **Mass Ref** | **Spin** | **Spin Ref** |
| Mrk 509 | 0.20 | [2] | $1.3 \times 10^8$ | [18] | — | — |
| NGC 4051 | 0.037 | [10] | $1.7 \times 10^6$ | [18] | — | — |
| NGC 1365 | 0.0167 | [43] | $2.0 \times 10^6$ | [44] | $> 0.97$ | [44] |
| Mrk 766 | 0.061 | [40] | $6.3 \times 10^6$ | [18] | 0.92 | [40] |
| MCG 6-30-15 | 0.0063 | [45] | $2.9 \times 10^6$ | [44] | 0.91 | [44] |
| Mrk 335 | 0.0167 | [46] | $2.5 \times 10^7$ | [47] | $> 0.99$ | [41] |
| NGC 4151 | 0.055 | [40] | $4.0 \times 10^7$ | [18] | 0.94 | [48] |
| IC 4329A | 0.098 | [10] | $5.0 \times 10^7$ | [18] | — | — |
| Mrk 279 | 0.22 | [40] | $2.5 \times 10^7$ | [18] | — | — |

| Mrk 79 | 0.092 | [10] | $4.0 \times 10^7$ | [18] | 0.7 | [49] |
|---|---|---|---|---|---|---|
| Mrk 841 | 0.055 | [10] | $1.6 \times 10^8$ | [18] | — | — |
| 1H0419-577 | 0.079 | [10] | $1.3 \times 10^8$ | [50] | — | — |
| Mrk 290 | 0.163 | [10] | $2.0 \times 10^7$ | [18] | — | — |
| MCG 5-23-16 | 0.116 | [10] | $5.0 \times 10^7$ | [18] | — | — |
| NGC 4507 | 0.119 | [10] | $6.3 \times 10^7$ | [18] | — | — |
| NGC 7582 | 0.285 | [10] | $5.0 \times 10^7$ | [18] | — | — |
| ESO 103-G35 | 0.056 | [40] | $2.5 \times 10^7$ | [18] | — | — |
| NGC 5506 | 0.246 | [40] | $1.6 \times 10^7$ | [18] | 0.93 | [51] |
| Swift J2127.4 + 5654 | 0.231 | [40] | $1.6 \times 10^7$ | [52] | — | — |
| IRAS 18325-5926 | 0.23 | [53] | $2.0 \times 10^6$ | [54] | — | — |
| PG 1448 + 273 | 0.15 | [55] | $2.0 \times 10^7$ | [55] | — | — |
| WKK 984438 | 0.319 | [56] | $2.0 \times 10^6$ | [56] | — | — |
| NGC 2992 | 0.21 | [42] | $1.0 \times 10^8$ | [18] | — | — |
| Zw I 1 | 0.265 | [57] | $3.2 \times 10^7$ | [58] | — | — |
| IRAS 13224-3809 | 0.173 | [59] | $1.0 \times 10^7$ | [60] | — | — |
| | | | **Jetted Nls1.** | | | |
| **Source** | $v/c$ | **Wind Ref** | $M_{BH}$ | **Mass Ref** | **Spin** | **Spin Ref** |
| 1H 0323 + 342 | 0.0028 | [24] | $2.0 \times 10^7$ | [61] | 0.8 | Landt + 2017 |

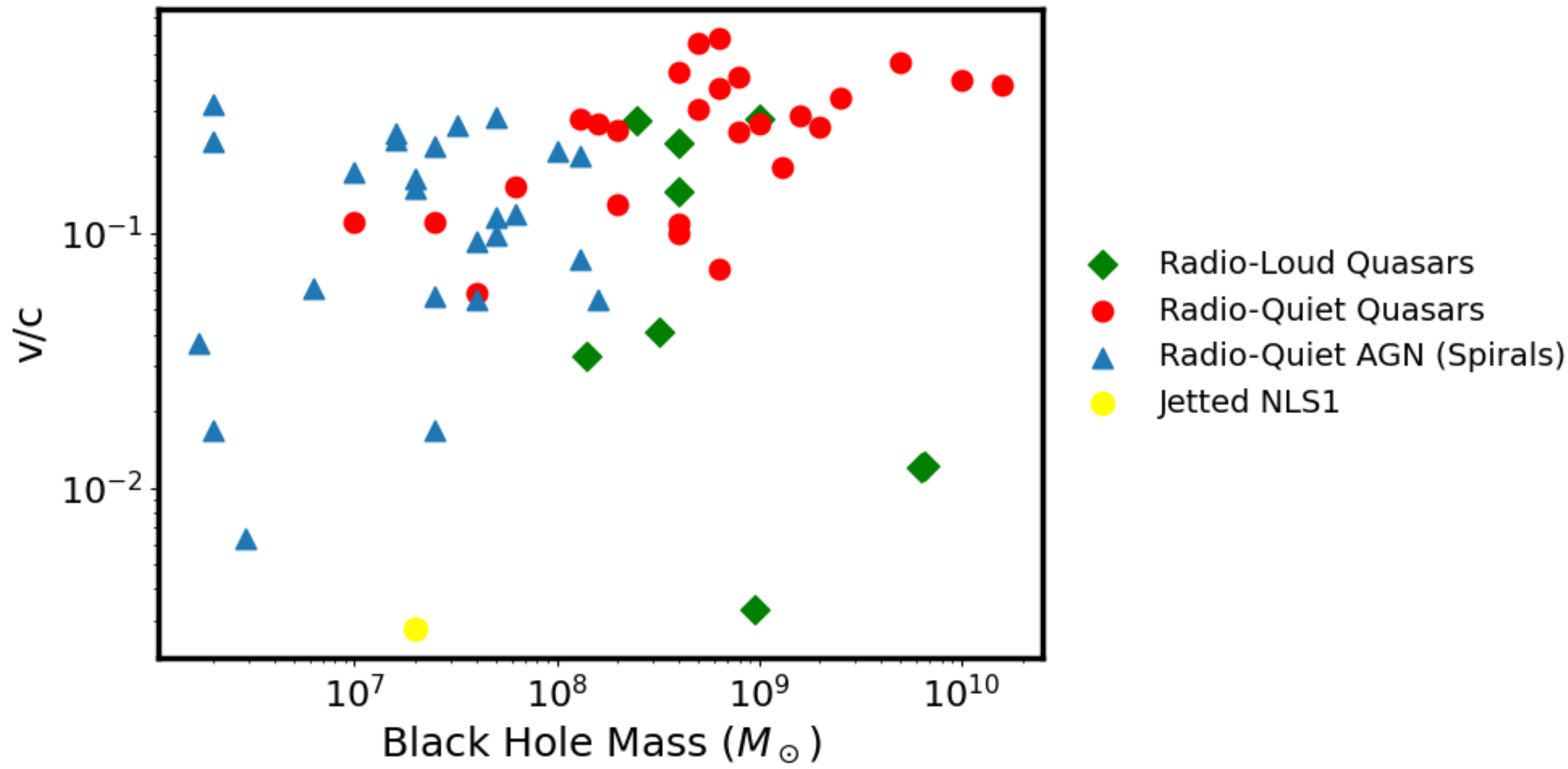

**Figure 1.** Outflow velocity (in v/c) from an ionized disk wind versus black hole mass for jetted and non-jetted AGN suggesting a stronger wind/jet tension in spiral AGN compared to elliptical AGN.

### *2.1. Radio Loud or Jetted Quasars*

The observations suggest that radio-loud quasars (RLQs) do not participate in the wind–jet anti-correlation as strongly as other AGN subclasses. While radio-quiet quasars (RQQs) define the highest-velocity wind population in the sample, with many systems exhibiting v/c $\gtrsim$ 0.3–0.5, RLQs are systematically shifted toward lower wind velocities despite hosting similarly massive black holes and powerful accretion flows. Importantly, RLQs do not exhibit the extreme wind suppression observed in jetted spiral systems such as 1H 0323 + 342, nor do they reach the highest wind velocities characteristic of RQQs. Instead, they occupy an intermediate regime in which moderately fast winds coexist with powerful relativistic jets. The data therefore indicate that the strongest form of the jet/wind anti-correlation operates primarily in radio-quiet systems, whereas RLQs appear to follow a distinct accretion regime in which jets and winds can coexist more readily.

In spiral-hosted AGN, where accretion flows are typically radiatively efficient and geometrically thin (e.g., [62]), the emergence of a jet appears to strongly inhibit the formation or acceleration of disk winds. The jetted NLS1 1H 0323 + 342 provides a clear ex-ample of this jet–wind anti-correlation: despite hosting a relativistic jet, its wind velocity is exceptionally low ($v/c \approx 0.0028$), placing it well below the radio-quiet Seyfert population at comparable masses. This behavior is consistent with a strong suppression of disk winds in systems dominated by cold, coherent accretion.

By contrast, RLQs—predominantly hosted in massive elliptical galaxies (e.g., [63])—do not follow this pattern. Several RLQs in our sample (e.g., PKS 1549–79, 3C 105, and MG J0414 + 0534) exhibit moderate to high wind velocities, overlapping with the lower end of the radio-quiet quasar distribution. Although some RLQs populate the low-velocity regime, the class as a whole does not display the systematic suppression seen in jetted spiral systems. Instead, the RLQ wind distribution is broad and continuous, indicating that jets and winds can coexist without strict mutual exclusivity. We argue that RLQs differ fundamentally from other subclasses of radio-quiet or jetless AGN in a way that weakens the degree of jet suppression.

Within this context, 3C 382 represents an important outlier. Its extremely low wind velocity ($v/c \approx 0.0033$) might at first appear to support strong wind suppression in RLQs. However, 3C 382 is not a clean, canonical high-excitation radio galaxy (HERG) and may instead occupy a transitional regime between efficient and inefficient accretion states which would tend to exclude it from our subclass as we will describe in the theory section. If so, its weak wind could reflect underlying changes in accretion properties (from HERG to LERG) rather than a generic consequence of jet production. This highlights the importance of accounting for subclass heterogeneity when interpreting the RLQ population.

The RLQ sample therefore demonstrates that the wind–jet anti-correlation is not a universal feature of black hole accretion, but rather a regime-dependent phenomenon. Strong wind suppression appears to be characteristic of jetted systems in spiral hosts, while in elliptical hosts, jets can coexist with moderately fast disk winds. This distinction underscores the importance of environmental and accretion-state differences in governing the relative dominance of outflows in active galactic nuclei.

### *2.2. Radio-Quiet or Non-Jetted Quasars*

The radio-quiet quasar (RQQ) population defines the high-velocity end of the wind distribution in our sample and provides the clearest evidence for a strong wind–jet anti-correlation in the most powerful AGN without jets. The measured outflow velocities span $\frac{v}{c} \sim 0.06$ to $v/c \sim 0.58$, with a mean value of $\left\langle \frac{v}{c} \right\rangle \approx 0.28$ and a median of $\tilde{v}/c \approx 0.27$. The close agreement between the mean and median indicates that the distribution is not dominated by outliers but instead reflects a genuinely high-velocity population. Indeed, a substantial fraction of RQQs exhibit $v/c \gtrsim 0.25$, and multiple sources reach $v/c \gtrsim 0.4$, placing them among the fastest disk winds observed in any AGN class.

In contrast to the RLQ population, the RQQ velocity distribution shows no evidence of suppression at high velocities. Rather, it extends smoothly to the largest values in the full sample, establishing RQQs as the upper envelope of wind speeds. This behavior is particularly significant given that these systems represent the most luminous, radiatively efficient AGN, yet lack powerful relativistic jets. The absence of jets in these high-accretion-rate systems appears to permit or facilitate the development of highly relativistic disk winds.

The RQQ sample spans nearly three orders of magnitude in black hole mass ($M_{\rm BH} \sim 10^7$–$10^{10}\ M_\odot$), but no statistically significant correlation is found between wind velocity and mass. High-velocity winds are observed across the entire mass range, indicating that mass is not the controlling parameter. Instead, the defining characteristic of the RQQ population is the simultaneous presence of high accretion power and the absence of large-scale jets.

Taken together, these results provide strong empirical support for a wind–jet anti-correlation in the most powerful AGN. While radio-loud systems do not exhibit a uniform suppression of winds, the highest-velocity outflows in the sample are found exclusively in radio-quiet quasars. This indicates that the most efficient production of relativistic disk winds occurs in systems where jet activity is absent, highlighting a fundamental division between wind-dominated and jet-dominated modes of black hole accretion.

*2.3. Non-Jetted Spiral Seyferts and Nls1*

The radio-quiet AGN hosted in spiral galaxies (Seyferts and NLS1s) exhibit wind properties that closely parallel those of radio-quiet quasars (RQQs), but at systematically lower black hole masses. The outflow velocities in this subclass span a broad range, from $\frac{v}{c} \sim 0.006$ to $v/c \sim 0.32$, with a mean value of $\left\langle\frac{v}{c}\right\rangle \approx 0.12$ and a median of $\tilde{v}/c \approx 0.09$. While these values are lower on average than those of RQQs, the distribution extends into the same high-velocity regime, with several sources exhibiting $\frac{v}{c} \gtrsim 0.2$ (e.g., NGC 7582, NGC 5506, IRAS 18325–5926, and WKK 984438).

The key distinction between this population and the RQQs lies not in the qualitative behavior of the winds, but in the characteristic mass scale. The spiral-hosted AGN span black hole masses of $M_{\mathrm{BH}} \sim 10^6$–$10^8\ M_\odot$, approximately two orders of magnitude lower than the typical RQQ mass range. Despite this, they are capable of launching moderately fast, and in some cases highly relativistic, winds. This demonstrates that the ability to produce strong disk winds is not restricted to the most massive black holes, but instead persists across a wide range of mass scales.

As in the RQQ population, no clear correlation is observed between wind velocity and black hole mass within the spiral sample. High-velocity outflows are found in both low-mass systems (e.g., WKK 984438 with $M_{\mathrm{BH}} \sim 2 \times 10^6\ M_\odot$ and $v/c \sim 0.32$) and more massive Seyferts, while relatively weak winds are also present across the same mass range. This lack of a monotonic trend reinforces the conclusion that black hole mass does not play a primary role in setting wind velocity.

Importantly, the spiral-hosted radio-quiet AGN define a continuous extension of the RQQ population toward lower black hole masses in the sense that both subclasses are capable of producing strong disk winds in the absence of powerful relativistic jets. However, this continuity does not imply identical spectral energy distributions (SEDs). The lower black hole masses characteristic of Seyferts naturally shift the thermal emission from the inner accretion disk toward higher characteristic frequencies, producing relatively stronger X-ray emission compared to the more UV/optically dominated SEDs of quasars. These differences arise naturally from accretion disk scaling with black hole mass and therefore affect the photoionization balance, ionization structure, and radiative coupling of the outflowing gas. Consequently, while radio-quiet quasars and radio-quiet Seyferts appear to occupy the same broad wind-dominated branch phenomenologically, the detailed ionization conditions and radiative transfer physics governing their winds are expected to differ systematically with black hole mass and associated SED properties.

This continuity further strengthens the evidence for a wind–jet dichotomy in AGN. While the most powerful winds are found in RQQs, the spiral-hosted systems demonstrate that the underlying physical mechanism responsible for these outflows operates efficiently even at much lower masses. The combined RQQ and spiral populations therefore define a coherent, wind-dominated branch of black hole accretion, distinct from the more heterogeneous behavior observed in radio-loud systems.

*2.4. The Jetted Nls1: 1H 0323 + 342*

The lone jetted NLS1 in the sample, 1H 0323 + 342, has

$$M_{\rm BH} \sim 2 \times 10^{7} M_{\odot}$$
$$v_{\rm out} \sim 0.003c.$$

This outflow velocity was derived in a model-dependent way by [24] and lies in the warm-absorber regime which is markedly lower than those of non-jetted NLS1 systems of comparable mass. This contrast is particularly informative. Its mass and Eddington ratios are comparable to Mrk 335 and Mrk 766. The key structural difference is the presence of a relativistic jet. The suppressed wind velocity in this object strengthens the emerging pattern. While only 1 jetted NLS1 has both a spin estimate, a black hole mass measurement, and an observed warm absorber, other similar systems such as Fairall 9 show similar features. But Fairall 9 and some other systems do not have estimated warm absorbers ([64]) so cannot be included in our statistics. Crucially, however, such objects show no strong UFO features, which contributes to the emerging pattern.

### *2.5. Statistical Summary*

Radio-quiet quasars (RQQs) exhibit systematically higher outflow velocities than both radio-loud quasars and radio-quiet spiral Seyfert/NLS1 systems. A Welch t-test comparing RQQs and radio-loud quasars yields $t \approx 3.1$ with $p \approx 0.004$, while a Mann–Whitney U test gives $p \approx 0.006$, indicating that the wind velocity distributions of the two populations differ significantly. The contrast between RQQs and radio-quiet Seyfert/NLS1 systems is similarly significant ($t \approx 3.8$, $p < 0.001$; Mann–Whitney $p \approx 0.002$), demonstrating that RQQs occupy a distinct high-velocity regime.

In contrast, radio-loud quasars and radio-quiet Seyfert/NLS1 systems do not show a statistically significant difference in their wind velocity distributions ($t \approx 0.6$, $p > 0.5$; Mann–Whitney $p > 0.4$). This overlap reflects the broader and more continuous velocity range of these two subclasses, in which moderately fast winds are present, but the highest-velocity outflows are absent.

Across the full sample, no statistically significant global correlation is detected between black hole mass and outflow velocity, despite the masses spanning nearly four orders of magnitude ($\sim 10^{6}$–$10^{10}\ M_{\odot}$). High-velocity winds are observed across a wide range of masses, while similarly massive systems can exhibit substantially weaker outflows, indicating that mass is not the controlling parameter.

Instead, the fastest winds ($v/c \gtrsim 0.4$) occur almost exclusively in radio-quiet quasars, while jet-hosting systems and lower-luminosity radio-quiet AGN are confined to more modest velocities. Notably, radio-loud quasars do not exhibit a uniform suppression of winds but rather occupy an intermediate regime overlapping with radio-quiet Seyfert systems. This behavior suggests that the wind–jet anti-correlation is strongest in the most luminous, jet-free systems and becomes weaker or absent in elliptical-hosted radio-loud quasars.

Taken together, these results argue against a purely mass-driven scaling for ultra-fast outflows and instead support a scenario in which additional parameters may provide the key triggering mechanism.

### *2.6. Physical Interpretation: Spin, Isco, and Disk-Black Hole Angular Momentum*

The observed wind–jet phenomenology cannot be explained by black hole mass and black hole spin alone. Instead, the decisive parameter appears to be the *direction* of the accretion disk angular momentum relative to the black hole angular momentum (i.e., aligned vs. anti-aligned). The relevant physics is encoded in the dependence of the innermost stable circular orbit (ISCO) on spin and relative angular momenta. For a Kerr black hole, the ISCO radius in gravitational units $r_g = GM/c^2$ is given by ([65])

$$r_{\mathrm{ISCO}} = r_g \left[3 + Z_2 \mp \sqrt{(3 - Z_1)(3 + Z_1 + 2Z_2)}\right],$$

where

$$Z_1 = 1 + (1 - a^2)^{1/3}\left[(1 + a)^{1/3} + (1 - a)^{1/3}\right],$$
$$Z_2 = \sqrt{3a^2 + Z_1^2},$$

$a$ is the dimensionless spin, and the minus sign corresponds to aligned angular momenta or prograde (co-rotating) orbits while the plus sign corresponds to anti-aligned angular momenta or retrograde (counter-rotating) orbits.

This dependence is highly asymmetric. For $a = +0.998$ (maximal prograde) → $r_{\mathrm{ISCO}} \approx 1.24\, r_g$, for $a = 0 \rightarrow r_{\mathrm{ISCO}} = 6\, r_g$, and for $a = -0.998$ (maximal retrograde) → $r_{\mathrm{ISCO}} \approx 9\, r_g$. The location of the ISCO determines the radiative efficiency

$$\eta = 1 - E_{\mathrm{ISCO}},$$

where $E_{\mathrm{ISCO}}$ is the specific binding energy at the ISCO. For prograde high spin, $\eta \sim 0.3$; for retrograde high spin, $\eta \sim 0.05$–$0.06$. Thus, prograde configurations concentrate gravitational energy release deep in the potential well, while retrograde disks terminate much farther out.

#### 2.6.1. Spiral Seyferts and Nls1: Co-rotation and Wind Dominance

In spiral galaxies, accretion is typically secular and long-lived ([62, 66, 67] promoting alignment between disk and black hole spin and high spin values. In a high-spin prograde configuration, the ISCO approaches the horizon, increasing radiative efficiency, disk temperature, and inner-disk radiation pressure. The enhanced energy density in the inner disk atmosphere naturally favors strong radiation-driven winds. Because the disk extends inward to $r_{\mathrm{ISCO}} \sim 1$–$2\, r_g$ for spin values above 0.94, the gravitational binding energy available to accelerate outflows is maximized.

In this regime, energy and magnetic flux are efficiently transported through the disk and its atmosphere, leaving comparatively less large-scale magnetic flux available to accumulate on the horizon for jet production ([68]). The result is a strong wind and suppressed jet formation. The jetted NLS1 1H 0323 + 342 illustrates this directly. While black hole spin (Figure 2 and Table 2) is notoriously subject to large uncertainties and model dependency, some work has shown that its black hole spin is likely lower than 0.8 ([61]), i.e., near the suppression threshold estimated at $a = 0.7$. Despite comparable mass to non-jetted Seyferts, its wind velocity is markedly lower, consistent with the conditions compatible with jet production. Thus, in co-rotating systems with intermediate black hole spin, the wind efficiency is less, and jets cannot be suppressed. Alternatively, in co-rotating, high-spin spiral systems $r_{\mathrm{ISCO}}$ goes down, $\eta$ goes up, wind strength goes up, and jet formation is disfavored.

**Table 2.** Jetted (1H 0323 + 342) and non-jetted AGN and black hole spin measurements.

| Source | $v/c$ | Spin $a$ |
|---|---|---|
| IRAS 13349 + 2438 | 0.27 | $> 0.93$ |
| ARK 120 | 0.269 | $> 0.85$ |
| NGC 1365 | 0.0167 | $> 0.97$ |
| Mrk 766 | 0.061 | 0.92 |
| MCG–6-30-15 | 0.0063 | 0.91 |
| Mrk 335 | 0.0167 | $> 0.99$ |
| NGC 4151 | 0.055 | 0.94 |
| Mrk 79 | 0.092 | 0.70 |
| NGC 5506 | 0.246 | 0.93 |
| 1H 0323 + 342 | 0.0028 | 0.80 |

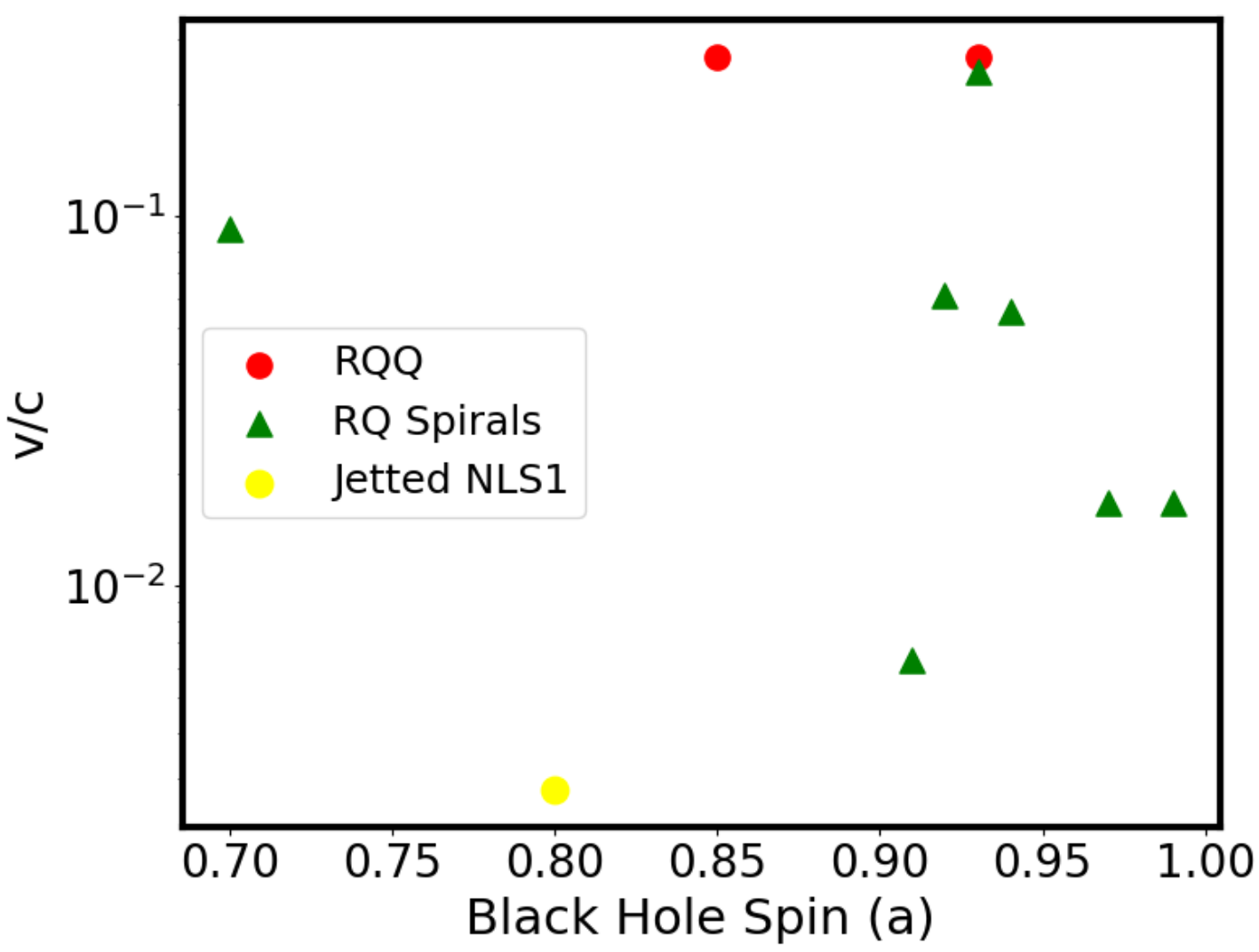

**Figure 2.** Outflow speed versus black hole spin. The prograde accretion model predicts that v/c should remain low below the threshold spin value and high above it in a black hole mass-scaled way.

On the other hand, Mrk 79 has a spin estimate of 0.7 ± 0.1 but is radio-quiet or jetless and its wind speed is high. If its actual spin value is $a > 0.7$, this may be an object that has recently suppressed its jet by way of a strong wind, i.e., the system has shifted from being similar to 1H 0323 + 342 and recently entered the radio-quiet subclass of AGN by crossing the threshold spin value. An increase in the number of sources with estimates of spin near the $a = 0.7$ value and wind outflow speeds, should be a focus of future observations.

2.6.2. FrII Quasars: Counter-Rotation and Jet Dominance

FRII quasars reside predominantly in elliptical galaxies where mergers are common. We do not revisit here the full dynamical spin evolution framework developed in previous work (starting with [69]). In that model, retrograde accretion episodes are expected to be rare and short-lived at high accretion rates (less than 10 million years), consistent with the observed minority fraction of powerful radio-loud quasars. The present work focuses instead on an independent observable—disk wind velocity—which provides a new diagnostic of the same underlying accretion configuration. While mergers do not typically produce retrograde accretion, they can generate misaligned episodes, that lead to counter-rotating phases. In that case, ISCO values are large, i.e.,

$$r_{\rm ISCO} \sim 9\, r_g (\text{for } a \approx 1),$$

placing the inner disk significantly farther from the black hole. This has a critical consequence. Lower radiative efficiency—the disk releases less gravitational energy in its inner regions, reducing the maximum acceleration available for radiation-driven winds. The absence of accretion close to the black hole horizon appears to lessen the wind-jet competition that exists in co-rotating high spin systems. Both jets and disk winds, thus, coexist, i.e., the wind is not *strongly suppressed* as in high spin, co-rotating systems. Because the disk truncates farther out, wind launching remains possible, but it never reaches the extreme velocities (~0.3–0.4c) seen in radio-quiet quasars with compact inner disks. Therefore, as $r_{\rm ISCO}$ goes up, $\eta$ goes down, making the wind strength moderate but spin still goes up so jet power is enhanced. Counter-rotation cannot produce the strong wind–jet anti-correlation observed in spirals because the disk never achieves the compact, high-efficiency configuration required to drive extreme winds.

More precisely, the Shakura-Sunyaev model ([70]) gives a disk dissipation function F(r) in terms of the ISCO

$$F(r) = \frac{3GM\dot{M}}{8\pi r^3}\left(1 - \sqrt{\frac{r_{\mathrm{ISCO}}}{r}}\right)$$

The radius at which the local dissipation reaches its maximum is obtained by differentiating $F(r)$ with respect to radius and setting the derivative equal to zero.
Writing

$$F(r) \propto r^{-3} - r_{\mathrm{ISCO}}^{1/2} r^{-7/2},$$

gives

$$-3r^{-4} + \frac{7}{2} r_{\mathrm{ISCO}}^{1/2} r^{-9/2} = 0.$$

Multiplying by $r^{9/2}$ yields

$$-3r^{1/2} + \frac{7}{2} r_{\mathrm{ISCO}}^{1/2} = 0,$$

so that

$$r^{1/2} = \frac{7}{6} r_{\mathrm{ISCO}}^{1/2},$$

or

$$r_{\mathrm{peak}} = \left(\frac{7}{6}\right)^2 r_{\mathrm{ISCO}} = \frac{49}{36} r_{\mathrm{ISCO}} \simeq 1.36\, r_{\mathrm{ISCO}}.$$

As $r_{\mathrm{ISCO}}$ decreases, the dissipation profile becomes increasingly concentrated toward smaller radii, leading to higher radiative efficiency and stronger inner-disk radiation fields.

#### 2.6.3. Two Distinct Regimes of Wind–Jet Interaction

The data therefore indicate two physically distinct modes

1. Co-rotating high-spin systems (spirals and most ellipticals):

- Small ISCO
- High efficiency
- Strong winds
- Jet suppression

2. Counter-rotating high-spin systems (a minority of ellipticals—i.e., FRII quasars):

- Large ISCO
- Moderate efficiency
- Weaker winds
- Powerful jets

A spin-only paradigm is insufficient because spin magnitude alone does not determine ISCO location; the direction of the disk angular momentum relative to the black hole angular momentum does. A maximally spinning prograde black hole and a maximally spinning retrograde black hole differ in ISCO radius by nearly an order of magnitude. Thus, we propose that the key governing parameter is not simply $|a|$, but the sign of $a$ relative to disk angular momentum (see Figure 3). The point we are making is not that prograde spin universally suppresses jets, but rather that within co-rotating systems there exists a gradual competition between winds and jets as spin increases. At intermediate prograde spin values, both outflows may coexist because the disk radiation field and associated wind strength are not yet sufficient to strongly inhibit jet production. As the black hole spins further up in the prograde direction, the ISCO moves closer to the horizon, increasing radiative efficiency and strengthening radiation-driven winds. Beyond some threshold spin value, which we tentatively associate with a~0.7, the wind becomes increasingly effective at suppressing the jet. In this sense, the transition is expected to be continuous rather than abrupt within the prograde regime itself.

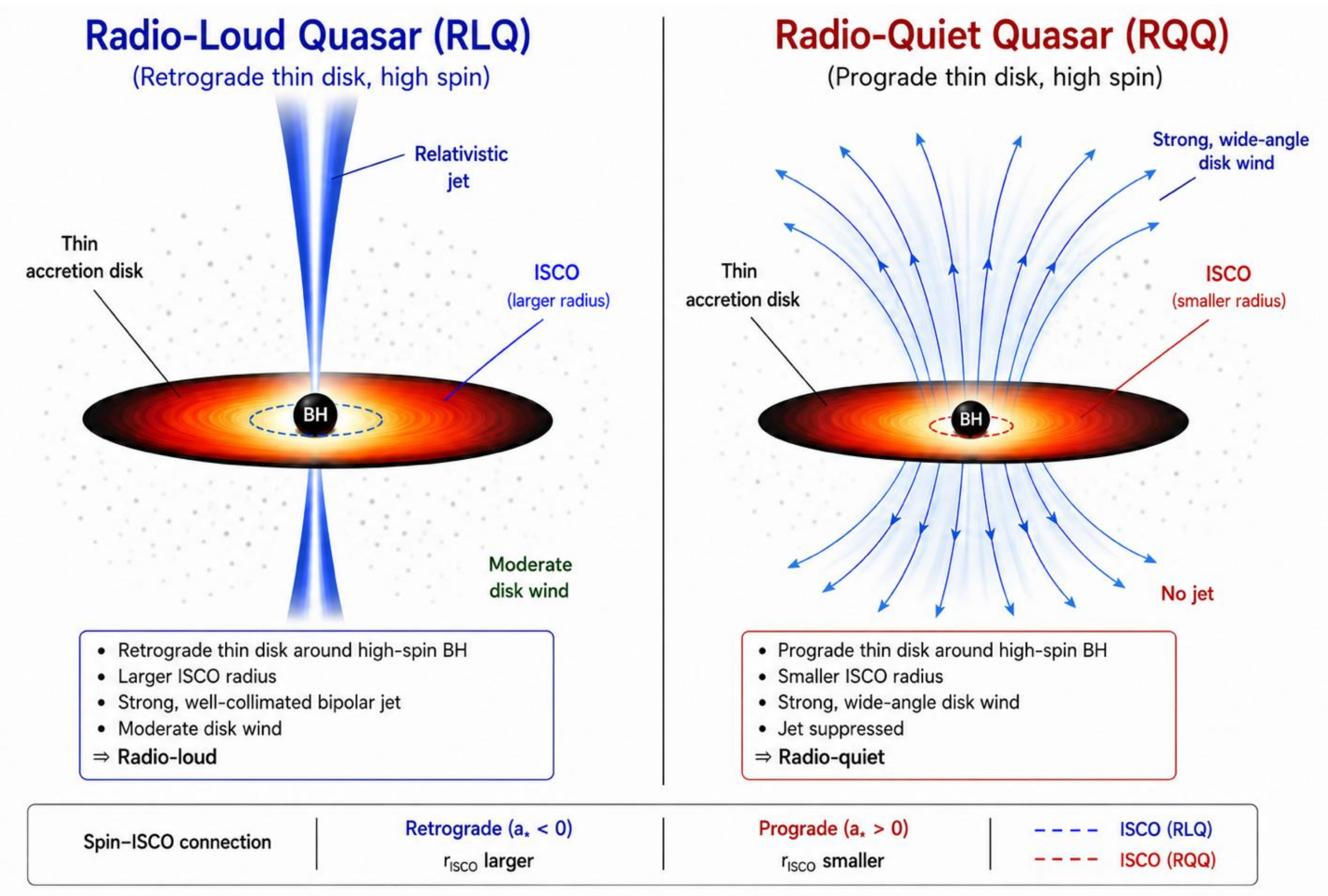

**Figure 3.** (**Right**) A co-rotating black hole with spin $a_* \geq 0.9$ producing strong winds from the inner disk that suppress the jet. (**Left**) A counter-rotating black hole with a strong jet and spin $a_* \geq 0.9$ producing weaker disk winds that fail to suppress the jet. Image AI generated.

The retrograde regime differs qualitatively because the larger ISCO prevents the inner disk from reaching the same radiative efficiency and compactness characteristic of high-spin prograde systems. As a result, even for large spin magnitudes, retrograde configurations produce only moderate winds while still permitting strong jet production. Thus, the distinction between prograde and retrograde accretion is not simply a degeneracy in spin magnitude, but reflects fundamentally different inner disk geometries and radiative environments.

It is important to also emphasize that ours is not a claim of pure scale invariance even for co-rotating accretion disks. This is because co-rotating disks in AGN cannot be treated in precisely the same way as co-rotating disks that undergo state transitions. In X-ray binaries, one plausible interpretation of the physics operating during the hard–to–soft transition is that the magnetic flux threading the black hole is not destroyed but reduced to some fraction $f$ of its hard-state value. If the hard-state magnetic flux is $\Phi_{\text{hard}}$, then in the soft state $\Phi_{\text{soft}} = f\Phi_{\text{hard}}$ with $f < 1$ (e.g.,[71] ). Since Blandford–Znajek jet power scales approximately as $P_{\text{jet}} \propto a^2\Phi_{\text{BH}}^2$ ([4]), the soft-state jet power becomes $P_{\text{jet,soft}} \propto a^2 f^2 \Phi_{\text{hard}}^2$. Thus, even a modest reduction in large-scale poloidal flux (e.g., $f \sim 0.1$) pro-duces a suppression of jet power by two orders of magnitude, naturally explaining the observed $10^2$–$10^3$ factor in radio quenching in soft states. High-spin systems may still produce detectable residual jets because the $a^2$ factor amplifies whatever flux remains. However, this simple "flux-fraction" scaling cannot directly apply to AGN. In co-rotating accretion disks in AGN—which we model for AGN in spiral galaxies as well as for the majority of merger induced accretion in ellipticals—a radiatively efficient disk will tend to remain in that state and the wind component will increase as the black hole spin increases ([72]). Assuming continuous accretion, the spin will eventually be high enough (possibly $a \sim 0.7$) for the wind to begin suppressing the jet. And even higher spin systems in these AGN will be radio-quiet or jetless, but with strong winds. The scale-invariant physics that applies to X-ray binaries and to spiral AGN, therefore, is that jets and winds are more strongly in opposition than for counter-rotating accretion systems. Any unified framework must account not only for geometric similarities in thin disks, but also for fundamentally different flux retention, accumulation timescales, or magnetic boundary conditions across the mass scale.

## 3. Conclusions

We have examined the relationship between accretion disk wind velocity and black hole mass across a diverse sample of active galactic nuclei, including radio-loud quasars, radio-quiet quasars, and radio-quiet AGN in spiral hosts. The results demonstrate that neither black hole mass, spin, nor the mere presence of accretion-driven activity is sufficient to explain the observed distribution of wind velocities. Instead, the strongest winds are found in radio-quiet quasars, while radio-loud quasars occupy an intermediate regime and jetted spiral systems exhibit the most suppressed outflows. We emphasize that our conclusions are not based on selectively chosen samples. Rather, we include all published measurements that satisfy our selection criteria. The number of available observations has now grown sufficiently to permit meaningful statistical comparisons across the AGN populations considered here.

Statistical tests confirm that radio-quiet quasars are distinct from both radio-loud quasars and radio-quiet Seyfert/NLS1 spiral systems, occupying a high-velocity regime characterized by $\frac{v}{c} \gtrsim 0.2$ and extending to $v/c \gtrsim 0.4$. In contrast, radio-loud quasars do not exhibit a uniform suppression of winds but instead overlap significantly with the Seyfert population, indicating a more complex relationship between jets and winds than a simple anti-correlation. The strongest version of the wind–jet anti-correlation is therefore confined to the most powerful, jet-free systems.

These findings present a challenge to the conventional paradigm in which powerful relativistic jets are primarily associated with high black hole spin. If high spin were the dominant factor, one would expect systems capable of launching strong jets to exhibit suppressed winds. Instead, the data show that the highest wind velocities are found in systems that are widely believed to host rapidly spinning black holes. This implies that high spin is not sufficient to guarantee jet production and, in fact, may be associated with conditions that favor strong disk winds leading to jet suppression.

One possible interpretation is that radio-loud quasars represent the high-spin analogs of radio-quiet quasars but in a different accretion state, analogous to the hard state in X-ray binaries where jets are prominent. However, this analogy does not hold. The hard state in X-ray binaries is radiatively inefficient, whereas both radio-loud and radio-quiet quasars are radiatively efficient systems. The true analogs of hard-state accretion in AGN are instead low-excitation radio galaxies (FRI LERGs), which are excluded from our sample. Therefore, differences in radiative efficiency or accretion state alone also cannot account for the dichotomy between radio-loud and radio-quiet quasars.

We propose instead that the key parameter distinguishing these systems is the relative angular momenta of the accretion flow and the black hole. In this framework, both radio-loud and radio-quiet quasars host rapidly spinning black holes and radiatively efficient accretion disks, but differ in whether the disk is co-rotating or counter-rotating with respect to the black hole. Co-rotating systems produce strong disk winds that can inhibit the formation of relativistic jets, leading to radio-quiet quasars. In contrast, counter-rotating configurations generate weaker winds, allowing efficient launching of powerful jets, as observed in radio-loud quasars.

This interpretation naturally explains the coexistence of high spin and strong winds in radio-quiet quasars, the absence of uniformly suppressed winds in radio-loud quasars, and the breakdown of simple wind–jet anti-correlation models. It further unifies the observed phenomenology across AGN subclasses without invoking fundamentally different central engines, instead attributing the diversity of outflows to differences in disk–black hole alignment.

Taken together, our results support a paradigm in which the relative dominance of winds and jets is governed not by black hole mass or spin alone, but by the geometric configuration of the accretion flow.

**Author Contributions:** Conceptualization, D.G., C.B.S. and M.N.; Formal analysis, A.M., M.B. and S.S; Investigation, A.M., M.B. and S.S; Writing – original draft, M.N.; Visualization, M.N. All authors have read and agreed to the published version of the manuscript.

**Funding:** This research received no external funding.

**Data Availability Statement:** The data are in the paper.

**Acknowledgments:** D.G. thanks Luigi Foschini and Stephanie Komossa for a discussion on a post by the former that got him to explore the topic that turned into this paper. We also thank two anonymous referees for detailed feedback.

**Conflicts of Interest:** The authors declare no conflict of interest.

## Abbreviations

| Abbreviation | Meaning |
|---|---|
| **AGN** | Active Galactic Nucleus (Nuclei) |
| **ADAF** | Advection-Dominated Accretion Flow |
| **BAL** | Broad Absorption Line |
| **BH** | Black Hole |
| **BLRG** | Broad-Line Radio Galaxy |
| **BZ** | Blandford–Znajek |
| **FRI** | Fanaroff–Riley Class I |
| **FRII** | Fanaroff–Riley Class II |
| **HERG** | High-Excitation Radio Galaxy |
| **ISCO** | Innermost Stable Circular Orbit |
| **LERG** | Low-Excitation Radio Galaxy |
| **MAD** | Magnetically Arrested Disk |
| **NLS1** | Narrow-Line Seyfert 1 |
| **RQ** | Radio-Quiet |
| **RQQ** | Radio-Quiet Quasar |
| **RL** | Radio-Loud |
| **RLQ** | Radio-Loud Quasar |
| **SMBH** | Supermassive Black Hole |
| **UFO** | Ultra-Fast Outflow |
| **UV** | Ultraviolet |
| **WA** | Warm Absorber |
| **XRB** | X-ray Binary |